\documentclass[]{gmaart2019}

\pdfoutput=1

\renewcommand{\varProcName}{Proc. 29. Workshop Computational Intelligence}
\renewcommand{\varLocation}{Dortmund}
\renewcommand{\varDate}{28.-29.11.2019}

\usepackage[utf8]{inputenc} 
\usepackage[T1]{fontenc} 
 
\usepackage{amsmath,amssymb,amsfonts,mathtools,amsthm} % extended math environments

\usepackage[babel,autostyle]{csquotes}

\usepackage{tabularx}
\newcolumntype{L}[1]{>{\raggedright\arraybackslash}p{#1}} % linksbündig mit Breitenangabe
\newcolumntype{C}[1]{>{\centering\arraybackslash}p{#1}} % zentriert mit Breitenangabe
\newcolumntype{R}[1]{>{\raggedleft\arraybackslash}p{#1}} 
\usepackage{booktabs}

\usepackage{paralist}   

\usepackage[noend]{algpseudocode}

\algrenewcommand\algorithmicrequire{\textbf{Voraussetzung:}}
\algrenewcommand\algorithmicensure{\textbf{Abschlussbedingung:}}

\usepackage{listings} % Alternative environment for algorithms
\usepackage{subcaption}  % To place figures side by side, if you wish to do that in latex... (package "subfigure" is deprecated)

\usepackage{epstopdf}
\usepackage{xcolor}
\selectcolormodel{gray}  % only use gray scale colors

\usepackage{scrhack}
\usepackage{varwidth}
\usepackage{rotating} % Rotate text (e.g. in tables)
\usepackage[defaultlines=2,all]{nowidow}  % no widow lines


\usepackage[version-1-compatibility]{siunitx} % Easily include SI units
\usepackage{verbatim}

\begin{document}

%\pagestyle{empty}
%\tableofcontents

\cleardoublepage
\setcounter{page}{1}
\pagestyle{scrheadings}

% At least two lines of text at the top and bottom of a page
\setnowidow[2]
\setnoclub[2]

% Include a single paper
%% Set title and author information
\renewcommand{\Title}{Data-driven charging strategies for grid-beneficial, customer-oriented and battery-preserving electric mobility}
% deprecated title: Towards grid-beneficial, battery-preserving and customer-oriented charging strategies in electric mobility
\renewcommand{\Authors}{Karl Schwenk\textsuperscript{1,2}, Tim Harr\textsuperscript{2}, René Großmann\textsuperscript{2}, Riccardo Remo Appino\textsuperscript{1}, Veit Hagenmeyer\textsuperscript{1}, Ralf Mikut\textsuperscript{1}}
\renewcommand{\Affiliations}{\textsuperscript{1} Institute for Automation and Applied Informatics,\\
	Karlsruhe Institute of Technology\\
	Karlsruhe, Germany\\
	E-Mail: karl.schwenk@kit.edu\\[1ex]
	\textsuperscript{2} eDrive Innovations, Daimler AG\\
	Sindelfingen, Germany\\
	E-Mail: karl.schwenk@daimler.com}
							 
%% Information required for the table of contents:
\renewcommand{\AuthorsTOC}{K. Schwenk, T. Harr, R. Großmann, R. R. Appino, V. Hagenmeyer, R. Mikut} % List (short) author names without superscripts and conjunction "and".
\renewcommand{\AffiliationsTOC}{Karlsruhe Institute of Technology, Daimler AG} % Required for the table of contents (only list names of universities / institutions)

%% Choose caption language of captions (english / german)!
%\setLanguageGermand
\setLanguageEnglish
							 
\setupPaper % generate title and toc

%\section*{Abstract}
%
\section{Introduction} \label{sec:intro}
\begin{figure}[h]
\centerline{\includegraphics[width=0.8\columnwidth]{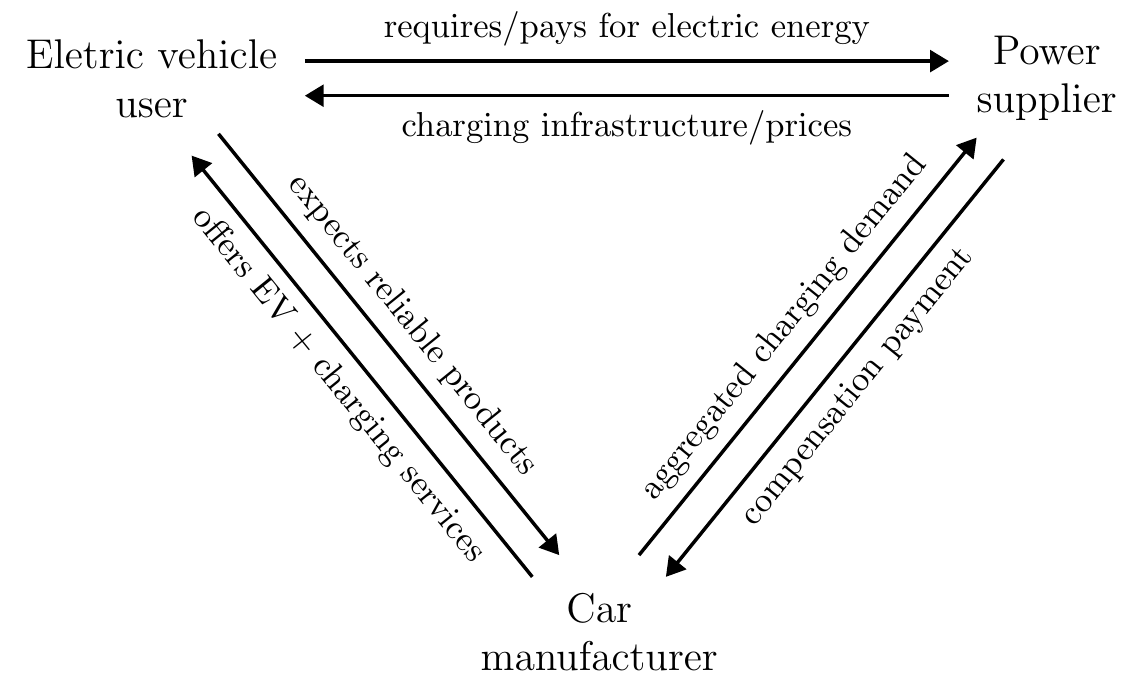}}
\caption{Interaction triangle of electric vehicle user, power supplier, and car manufacturer.}
\label{fig:trilemma}
\vspace{-0.3cm}
\end{figure}

%%%%%%%%%%%%%%%%%%%%%%%%%%%%%%%%%%%%%%%%%%%%%%%%%%%%%
% GENERAL
%%%%%%%%%%%%%%%%%%%%%%%%%%%%%%%%%%%%%%%%%%%%%%%%%%%%%
%
% Situation \cite{IEA2017}
Electric Vehicle (EV) penetration, renewable energies, and customer orientation of car manufacturers \cite{IEA2018} enables synergies between energy supply, vehicle users, and the mobility sector. 
However, also new challenges arise \cite{Eider2017} which we target to examine with three types of agents and their perspectives: EV user, power supplier and car manufacturer.
Although many research papers in the literature describe single perspectives, a threefold contemplation of their connections as shown in Fig. \ref{fig:trilemma}, is still missing. 
In the following, we briefly present the single perspectives and respective interactions.
To satisfy their individual needs, \textbf{vehicle users} expect their EVs to be just as reliable and convenient to use, as known from combustion engine cars \cite{Broadbent2019, Luetzenberger2015}.
However, limited range and longer charging times of EVs complicate individual mobility, manifesting e.g. in \textit{range anxiety} \cite{Eider2017, Cuchy2018}.
%
% INTERACTIONS
For EV users also the power supply interaction changes, as not only home appliances but also mobility requires electric energy.
Hence, a convenient and cost-efficient charging process is hard to achieve without automation technology, when using an EV on a daily basis \cite{Schwenk2019}.

%\subsection{Power supplier}
%%%%%%%%%%%%%%%%%%%%%%%%%%%%%%%%%%%%%%%%%%%%%%%%%%%%%
% GRID
%%%%%%%%%%%%%%%%%%%%%%%%%%%%%%%%%%%%%%%%%%%%%%%%%%%%%
%
% INTERESTS & CHALLENGES
\textbf{Power suppliers} (distribution grid operators, energy retailers) target an efficient and fail-proof grid operation, for which a reliable forecast and control of electric loads is desired \cite{Liu2015, Tan2016}.
However, the intermittent nature of renewable energy production endangers the utility grid through voltage and frequency fluctuation \cite{Allison2018, Aliasghari2018}.
The energy demand caused by charging EVs might further amplify this effect \cite{Jarvis2019}.
%
% INTERACTIONS 
To avoid these issues, power suppliers may influence the EV users' affection to connect their vehicles via dynamic energy prices \cite{Allison2018, Aliasghari2018, Alipour2017}.
At the same time, multiple EVs connected to the grid---especially if bi-directional charging is enabled \cite{Mozafar2017}---help power suppliers to level out imbalances due to renewable energy generation \cite{Appino2019}.

%\subsection{Car Manufacturer}
%%%%%%%%%%%%%%%%%%%%%%%%%%%%%%%%%%%%%%%%%%%%%%%%%%%%%
% MANUFACTURER
%%%%%%%%%%%%%%%%%%%%%%%%%%%%%%%%%%%%%%%%%%%%%%%%%%%%%
%
% INTERESTS & CHALLENGES
%Novel mobility concepts \cite{Smith2018}, e.g. car sharing, challenge the hitherto business model of \textbf{car manufacturers}, i.e. producing and selling cars.
Electric mobility also poses new issues for \textbf{car manufacturers}:
During charging and discharging of EV batteries a degradation (\textit{battery aging}) occurs \cite{Maia2019}, that correlates with a value depreciation of the entire EV.
%
% INTERACTIONS
However, EV users' satisfaction requires reliable and value-stable products, which car manufacturers aim to achieve by offering services, such as charging assistants \cite{Schwenk2019}.
The provided charging strategies target simplified and sustainable EV usage by considering individual customer needs and battery aging.

%%%%%%%%%%%%%%%%%%%%%%%%%%%%%%%%%%%%%%%%%%%%%%%%%%%%%
% TO COME
%%%%%%%%%%%%%%%%%%%%%%%%%%%%%%%%%%%%%%%%%%%%%%%%%%%%%
%
The remainder of this paper is structured as follows: 
To identify and develop missing models of the outlined problem, a general approach is presented in Section \ref{sec:approach}.
Then, Section \ref{sec:framework} describes an online learning framework for data acquisition, storage and application.
In Section \ref{sec:estimation}, we propose two data-driven consumption models.
Finally, Section \ref{sec:conclusion} gives a brief summary and outlook on future work.
\section{Approach} \label{sec:approach}
\begin{figure}[h]
	\centerline{\includegraphics[trim = 0mm 0mm 0mm 0mm, clip, width=1\columnwidth]{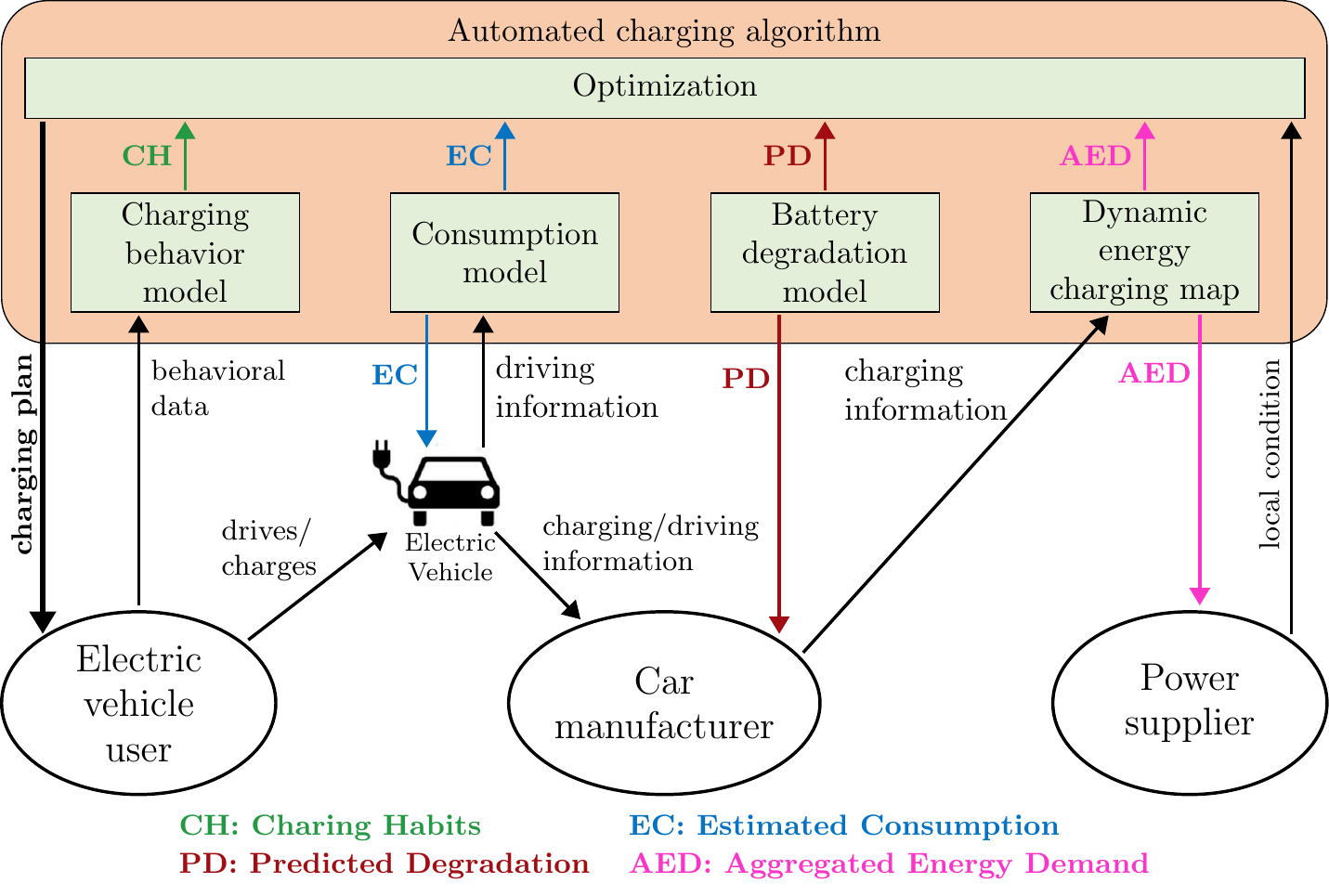}}
	\caption{Schematic map of models to describe perspectives and interactions, Automated charging algorithm (orange box) comprises missing models (green boxes)}
	\label{fig:model_flow}
	%\vspace{-0.3cm}
\end{figure}

Despite existing model predictive approaches \cite{Giorgio2014, Rajaei2019} that support EV users while charging, individualized charging strategies are in general still inadequately dealt with in the literature.
We target to identify missing models that quantify perspectives and interactions, for which Fig. \ref{fig:model_flow} shows a schematic map (the green boxes represent unexplored models requiring further evaluation).

To automatically create charging plans that satisfy the interests of power suppliers, EV users, and car manufacturers at the same time (cf. Section \ref{sec:intro}), an \textbf{Automated charging algorithm} will be required (Fig. \ref{fig:model_flow}, orange box).
A \textbf{Charging behavior model} (cf. Section \ref{subsec:behavior_model}) could determine Charging Habits (CH, Fig. \ref{fig:model_flow}, green arrow) based on behavioral data of \textbf{Electric vehicle users}.
Thereby, charging plans provided to the EV user can be adapted to individual requirements.
In order to obtain user- and route-specific Estimated Consumption (EC, Fig. \ref{fig:model_flow}, blue arrows) for the EV, a data-driven \textbf{Consumption model} (cf. Section \ref{sec:estimation}) may use driving information of the EV.
The EC can also help to estimate the remaining driving range of the EV.
Further, charging plans should be adapted to be battery preservative by means of a \textbf{Battery degradation model} (cf. Section \ref{subsec:battery_model}) that provides a Predicted Degradation (PD, Fig. \ref{fig:model_flow}, red arrows) for specific charging strategies.
Using the PD, car manufacturers may also monitor the condition of operating EV batteries.
A \textbf{Dynamic charging energy map} (cf. Section \ref{subsec:energy_map}) could calculate the Aggregated Energy Demand (AED, Fig. \ref{fig:model_flow}, pink  arrows) based on charging information of several EVs.
The \textbf{Power supplier} may then make use of the AED for improved load forecast precision.
Together with information about momentary grid load conditions obtained from Power suppliers, peak shaving and load balancing applications can be included in the charging plan.
An \textbf{Optimization} approach may finally conclude all the information and calculate a charging plan for the EV user.
\subsection{Charging behavior model} \label{subsec:behavior_model}
The effectiveness of charging strategies depends on user acceptance, which in turn requires user-specific charging strategies \cite{Schwenk2019}.
Therefore, a suitable characterization of users' driving and charging behavior is necessary \cite{Kuehl2019}.
Existing approaches in the literature distinguish between qualitative and quantitative ones.
Psychological approaches, e.g. deducted from customer surveys \cite{Azadfar2015, Eider2017, Lueddecke2016} allow to characterize basic user types.
Therein, \textit{gamification} and \textit{incentivation} \cite{Eider2017} methods are developed to increase user acceptance.
%The obtained knowledge is a valuable starting point for further evaluations.
Quantitative approaches aim at predicting driving and charging behavior of EV users, e.g. to infer energy demand \cite{Desai2018, Li2019a, Dong2016, Sun2019}. 
However, an adequate user integration, i.e. individualized charging strategies based on EV users' charging habits is not represented.

To attain a more universal characterization of the EV users' charging behavior, different user types need to be determined by analyzing behavioral data of the EV user.
Thereby, different user clusters can be identified, for which individual charging habits can be determined (cf. Fig. \ref{fig:model_flow}, left).
To clarify the EV users' objectives, we plan a customer survey in future work.
Therein, the basic requirements towards automated charging strategies are inquired from 
i) customers using a conventional car,
ii) customers about to acquire an EV, 
iii) customers just recently started using an EV, and
iv) customers already using an EV for a longer time.
Specific questions on the users' mobility requirements, charging habits, doubts, and expectations are supposed to reveal further requirements for individualized charging strategies.
Subsequently, a data-driven analysis of charging processes and EV user information may support these findings and may allow creating a mathematical model.
\subsection{Battery degradation model} \label{subsec:battery_model}
The interest of car manufacturers focuses on the depreciation process of the EV and its key factors.
As the battery considerably contributes to the vehicle value \cite{Neubauer2012}, we particularly consider battery degradation. 
Present-day mobility concepts (e.g. car sharing, leasing) comprise the vehicle battery to remain property of the manufacturer \cite{Smith2018}.
Value-stable batteries are thus even more relevant to operate economically.
To quantify battery aging, usually the State of Health\footnote{Ratio of actual usable capacity in relation to the nominal battery capacity} (SOH) is used \cite{Lee2015, Zheng2019}.
Hitherto SOH models have limited practicability due to a complex execution \cite{Eddahech2012} or inaccurate State of Charge (SOC) estimations \cite{Hannan2017}.
Data-driven methods, e.g. Bayesian networks and neural networks \cite{Susilo2018, Li2019} hold feasible alternatives for SOH estimation \cite{Berecibar2016}.
However, the existing approaches barely use user-related data, for which reason the influence of charging behavior on battery aging is not represented properly.

Hence, we target to evaluate the influence of charging strategies on battery degradation (cf. Fig. \ref{fig:model_flow}, center right).
By means of a neural network regression model, we target to estimate the energy consumption for driving (cf. Section \ref{sec:estimation}), not taking any battery aging influences into account.
The model trained on data from batteries without degradation can be used to estimate the energy consumption for EVs with aged batteries.
A discrepancy between the estimation and the real consumption indicates a battery aging caused by increased internal losses.
Together with a feature-augmented map, this model can additionally estimate energy consumption for charging scheduling, that also considers individual driving styles and environmental conditions. 
Once this battery aging estimation is mature, also a dependency on charging influences could be inferred.
A concept validation has to be proceeded with both simulation and real data.
\subsection{Dynamic charging energy map} \label{subsec:energy_map}
To match energy usage and generation, power suppliers require information on the energy demand at a certain time and place, targeting a fault-free delivery of electric energy. 
This includes the demand due to charging EVs.
Research papers to model the energy demand caused by charging EVs already exist.
They differentiate among approaches to predict and control the grid condition \cite{Jarvis2019, Sundstrom2010, Majidpour2016, Appino2019}, and approaches for dimensioning and locating new charging stations \cite{Gonzalez2014, Li2018}.
Stochastic assumptions to describe the EV user behavior are a major deficiency of these approaches.

By combining information about operating EVs, the behavior of their drivers, and the associated charging requirements, we target to aggregate the energy demand caused by charging vehicles and predict vehicles' anticipated charging location.\footnote{The access to EV information subjects to the present data security legislation \cite{DSGVO2018}.}
EVs requiring charging can be consolidated to larger energy demands characterized by a load profile over time and location.
Then, an ``energy map'' can be created, containing the aggregated energy demand in real time, or a prediction for future points in time (cf. Fig. \ref{fig:model_flow}, right).
The required energy can either be acquired directly from the energy market, or the information can be passed to power suppliers to handle the predicted load accordingly.
For concept validation, a simulation environment has to be developed.
\section{Data framework} \label{sec:framework}
\begin{figure}[h]
	\centerline{\includegraphics[trim = 0mm  0mm 0mm 0mm, clip, width=1\columnwidth]{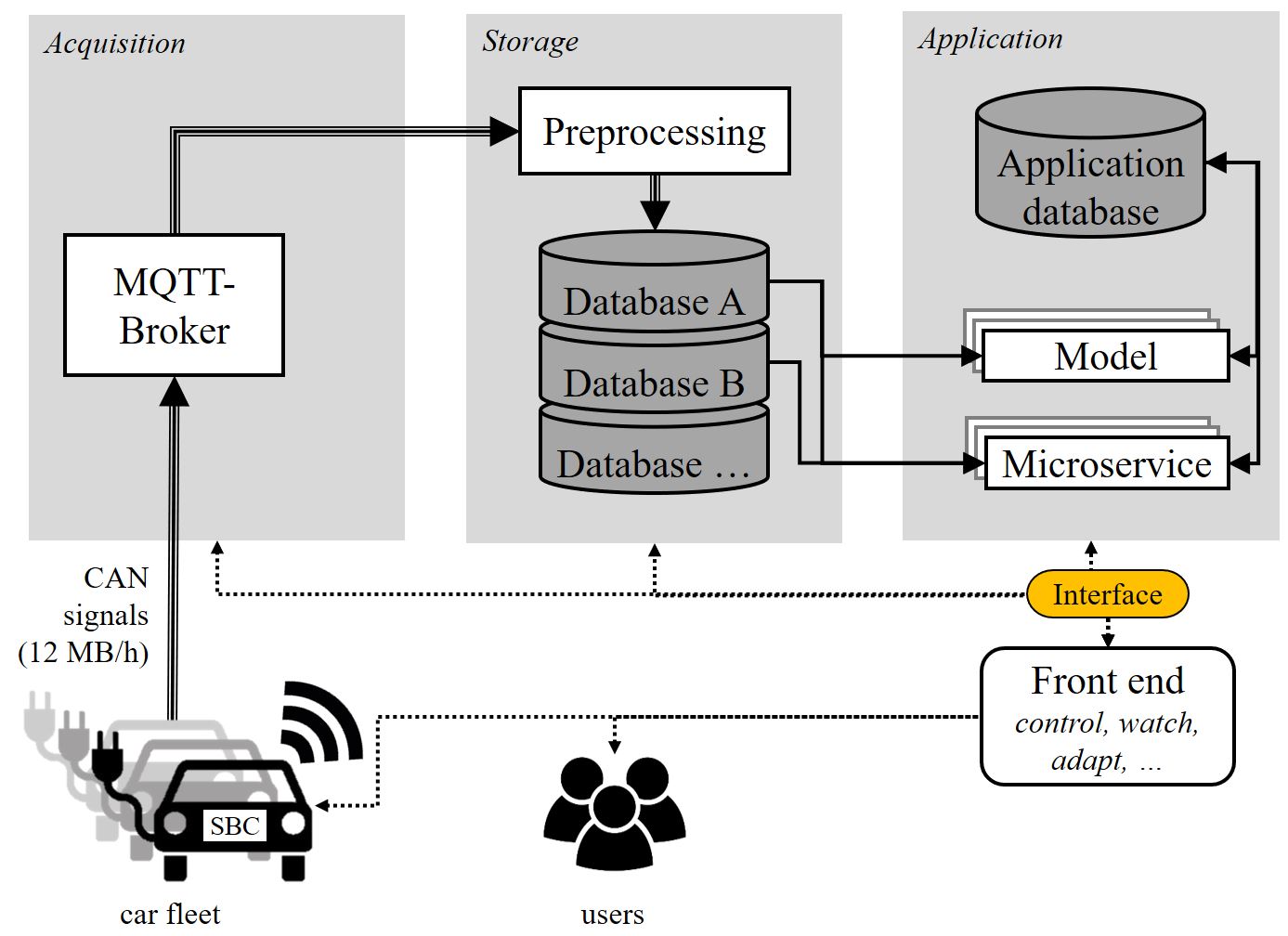}}
	\caption{Scheme of framework for data acquisition via a MQTT-broker, storage in several databases and application of data-driven models or microservices.}
	\label{fig:data_framework}
	%\vspace{-0.3cm}
\end{figure}

In the following, we propose a cloud-based framework to acquire, store and analyze vehicle data, as shown in Fig. \ref{fig:data_framework}.
A telematic system records selected Controller Area Network (CAN) signals for a fleet of connected EVs during driving and charging. 

% data acquisition
Applications, e.g. distributed functions structured in microservices require a reliable internet connection to provide historical and live data.
Existing devices such as data loggers store data in the car locally.
However, data can only be transmitted infrequently via infrastructure-bound connections (e.g. WLAN or LAN).
A real-time vehicle status can thus not be established and the large amount of data causes the data transmission to be time-intensive. 
For this reason, we use a single-board computer (SBC) equipped with a GPS antenna and mobile internet connection as a gateway between vehicle and back-end infrastructure.
The SBC initially decodes and filters data with libraries stored on its internal memory.
Via software packages, the SBC can be updated and controlled remotely not requiring a physical contact to the SBC.
The data transmission utilizes a Message Queuing Telemetry Transport (MQTT) protocol following a publish and subscribe mechanism\footnote{participants can publish information others can subscribe to}.
A central MQTT broker (Fig. \ref{fig:data_framework}, left) organizes the data distribution.
This mechanism also enables communication into the vehicle.
Currently ten EVs provide 370 signals of interest, which we record with a 10-Hz sampling rate\footnote{equals data stream of approx. 12 Megabyte per hour and vehicle.}. 

% data storage
We process and store all collected data according to its structure (Fig. \ref{fig:data_framework}, middle).
A relational database stores vehicle data and meta-information.
Further, a document-oriented database allows to store semi-structured signal time series data efficiently.
Additionally, a library database allows transforming vehicle specific data into a standardized format.

% application
The stored data can subsequently be used to build models or provide stand-alone microservices with information (Fig. \ref{fig:data_framework}, right).
Each model or microservice can make use of several data sources, or communicate among each other, respectively.
A separate application database contains configurations and meta-information necessary for the microservice operation.
The cloud structure allows continuity in various aspects.
Models can be easily implemented, changed, monitored and deleted.
Software changes can be applied and deployed continuously. 
Through a constant availability, the models can be fed with new data each time it is produced.
This allows a continuous development of the model with a updated data stream. 
However, a continuous availability also constitutes a risk in terms of information security \cite{Chen2012}, that needs to be handled accordingly. 
The modular structure allows to create and train several models in parallel.
An Application Database contains all models and their internal, learned parameters (Fig. \ref{fig:data_framework}, right).
Via a human machine interface, the models and their learning progress can be monitored and analyzed.
This allows to supervise the data acquisition, storage, and application.
Furthermore, databases can be examined, microservices and models can be adapted, and errors can be handled accordingly.
The interface also provides feedback to the vehicle fleet and vehicle users.
\section{Data-driven consumption model} \label{sec:estimation}
Here, we propose a data-driven consumption model as an exemplary application of the framework presented in Section \ref{fig:data_framework}.
Consumption models, as used in charging scheduling, help to estimate energy demands for routes the user will drive \cite{Baum2019}.
However, existing estimators mostly utilize physical models neglecting influences of the driver and the environment.
Such consumption models yield relative estimation errors between 2.52 \% and 8.3 \% \cite{Fiori2016, Hong2016}.
A higher model accuracy allows to compute a more reliable and thoroughly adapted charging strategy.
Thus, we aim to design a model estimating the total consumed energy per driven distance for a given driving behavior, environmental condition, and battery state.
Note, that we use the dimensionless State of Charge (SOC) as a metric for the consumed energy. 
\subsection{Data selection} \label{subsec:data_selection}
To abstract driving and environmental influences we use time series data recorded during driving of the EVs.
We differentiate the input signals in the three categories: driving behavior, battery state, and environmental condition (as shown in Table \ref{tab:signal_selection}).
The \textit{acceleration torque}, \textit{brake torque}, and \textit{recuperation torque} are measured for front and rear axle, i.e. two signals for those values exist.
Thus, we obtain a total signal number of $N_\textrm{signal}=15$.
\begin{table}[h]
\begin{footnotesize}
  \begin{center}
  \caption{Input signals grouped by driving behavior, battery state and environmental condition}
  \begin{tabular}{lll}
	\toprule
	    \textbf{Driving behavior}& \textbf{Battery state} & \textbf{Environmental condition} \\
    \midrule
		acceleration torque & total electric current & geodetic altitude \\
		brake torque & battery voltage & ambient temperature \\
 		recuperation torque& battery temperature & \\
		lateral acceleration & &\\
		vehicle speed & & \\		
		electric current driving & & \\	
		longitudinal acceleration  & & \\	
    \bottomrule
  \end{tabular}
  \label{tab:signal_selection}
  \end{center}
\end{footnotesize}
\end{table}

With a variety of driving situations and environmental conditions as input, we target a universal estimator that generalizes well on arbitrary input data.
The samples are sections with duration $t_\textrm{sec}=6\textrm{ min}$ of EV trips representing a typical ride.  
All trips with insufficient duration or faulty signal values are discarded beforehand.\footnote{In following evaluations, these special cases need to be contemplated separately.}
To avoid estimation errors due to random noise, the originally 10-Hz-sampled signals are aggregated before passing them to the model.
For each signal and for all values within an aggregation time period $t_\textrm{agg}=1\textrm{ min}$ the mean is calculated.

With $15$ signals multiplied by $6$ data points per trip section we obtain the extracted features $x_{1,..,90}$.
Note that we chose the parameters $t_\textrm{sec}$ and $t_\textrm{agg}$ according to preliminary test results.\footnote{In future work, a more detailed evaluation on the selection of these parameters is required.}

For each data sample representing a trip section of $t_\textrm{sec}=6\textrm{ min}$ we calculate a label 
\begin{align}
	\Gamma=\dfrac{\overline{e} - \underline{e}}{\overline{o} - \underline{o}}.
\end{align}
Therein, $\underline{e}$ is the SOC at the start of the trip section, $\overline{e}$ the SOC at the end.
Similarly, $\underline{o}$ is the mileage (in kilometers) at the start of the trip section, $\overline{o}$ at the end, respectively.
%%
%\begin{figure}
%	\centerline{\includegraphics[trim = 0mm 0mm 0mm 0mm, clip, width=1\columnwidth]{density_gamma}}
%	\caption{Density distribution of the training data evaluated for the specific consumption $\Gamma$}
%	\label{fig:density_gamma}
%	%\vspace{-0.3cm}
%\end{figure}
%%%
%
%
\subsection{Models}
To model the dependency between the input features $x_{1,..,90}$ and output label $\Gamma$, we design two models.

\subsubsection{Model A}
For the first Model A we use a linear regression model
\begin{align}
	\hat{\Gamma}=\mathbf{W}\cdot\mathbf{x}+b,
\end{align}
with the input features $\mathbf{x}=(x_1,x_2,...,x_{90})^\top$ of the aggregated signal time series, the weight matrix $\mathbf{W}\in\mathbb{R}^{1\times90}$, and the bias $b\in\mathbb{R}$.
The weights $\mathbf{W}$ and the bias $b$ are chosen to yield a minimum mean squared error (cf. Section \ref{subsec:evaluation}) between the training samples and the regression \cite{Weisberg2005}.

\subsubsection{Model B}
\begin{figure}[h]
	\centerline{\includegraphics[trim = 0mm 0mm 0mm 0mm, clip, width=1\columnwidth]{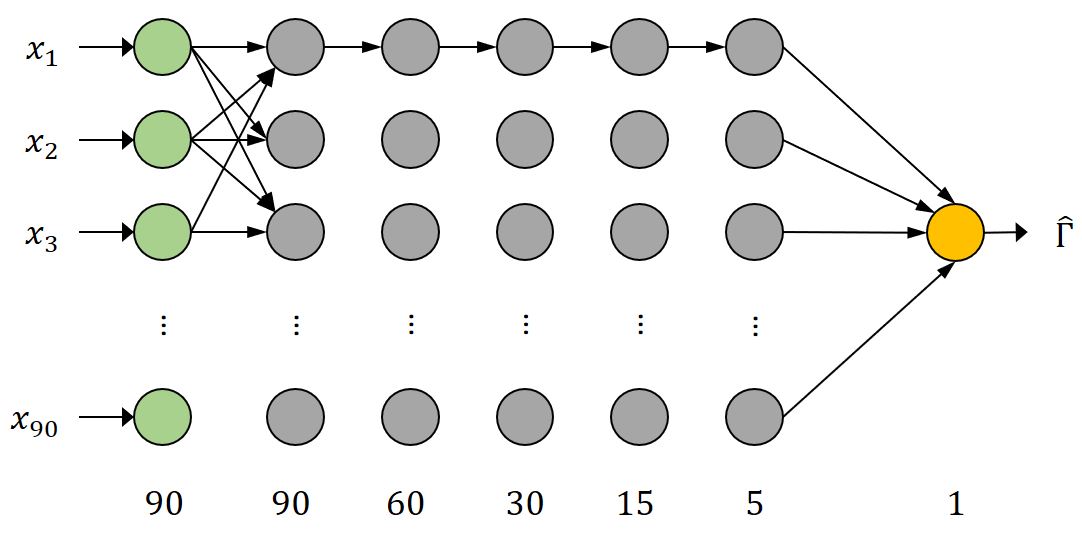}}
	\caption{Schematic architecture Model B, green: input nodes, hidden layer nodes in grey, output node in orange, total number of nodes below each layer.}
	\label{fig:neural_net}
	%\vspace{-0.3cm}
\end{figure}

For the second Model B we design a neural network regression model.
The input layer comprise 90 nodes representing the input features $x_{1,..,90}$.
Further, we use five hidden layers in a triangular shape, as shown in Fig. \ref{fig:neural_net}.
Each of the hidden layer nodes is activated through a Rectified Linear Unit (ReLU).
The output layer of the neural network consists of one node representing the estimated consumption $\hat{\Gamma}$.

For training the model we process all trips into training samples (cf. Section \ref{subsec:data_selection}).
Then, the data is fed to the models in batches of 32 samples over 100 epochs.
The model is implemented in \textit{Python} \cite{Python2019} using \textit{Keras} \cite{Keras2015}.
We use the \textit{Adam} optimizer \cite{Kingma2014} with a learning rate $\alpha=0.001$.
\subsection{Evaluation}\label{subsec:evaluation}
The models are trained with data obtained from eight out of ten EVs.
For validation, the data obtained from the remaining two EVs is used, i.e. we compare the estimated consumption $\hat{\Gamma}$ with the actual consumption $\Gamma$.
To quantify the model performance, we calculate the metrics\\
Mean Absolute Error (MAE)
\begin{align} \label{eq:mae}
	\textrm{MAE} = \dfrac{1}{N}\sum_{n=1}^N |\Gamma_n-\hat{\Gamma}_n|,
\end{align}
Relative Mean Absolute Error (RMAE)
\begin{align} \label{eq:rmae}
	\textrm{RMAE} = \dfrac{1}{N}\sum_{n=1}^N |\dfrac{\Gamma_n-\hat{\Gamma}_n}{\Gamma_n}|,
\end{align}
Mean Squared Error (MSE) 
\begin{align} \label{eq:mse}
	\textrm{MSE} = \dfrac{1}{N}\sum_{n=1}^N (\Gamma_n-\hat{\Gamma}_n)^2,
\end{align}
and Root Mean Squared Error (RMSE)
\begin{align} \label{eq:rmse}
	\textrm{RMSE} = \sqrt{\dfrac{1}{N}\sum_{n=1}^N (\Gamma_n-\hat{\Gamma}_n)^2}.
\end{align}

Table \ref{tab:errors} reports the validation results according to the error metrics \eqref{eq:mae}-\eqref{eq:rmse}.
The results show, that the neural network Model B outperforms the linear Model A in all four metrics. 
Using Model B, the estimation yields a mean absolute error of $0.00461\textrm{ km}^{-1}$, i.e. for a trip of $1\textrm{ km}$ driven distance, the battery level change is estimated with an average discrepancy of $0.461$\% SOC compared to the true value.
On the contrary, Model A estimates the battery level change for such a trip with an average discrepancy of $52.783$\% SOC.
Note that an SOC of $100$\% represents a fully charged battery and an SOC of $0$\% an empty battery, respectively.
Considering this fact, the estimations obtained from Model A do not provide useful information on the EV's consumption. 

\begin{table}[h]
\begin{footnotesize}
  \begin{center}
  \caption{Evaluation of Model A and Model B with MAE, RMAE, MSE, and RMSE. }
  \begin{tabular}{lcccc}
	\toprule
	    \textbf{Model} &  \textbf{MAE [$\textrm{km}^{-1}$]} & \textbf{RMAE [-]} & \textbf{MSE [$\textrm{km}^{-2}$]} & \textbf{RMSE [$\textrm{km}^{-1}$]} \\
    \midrule
    	\textbf{Model A} & 0.52783 & 1.45031 & 0.44548 & 0.66744 \\ 
	   	\textbf{Model B} & 0.00461 & 0.01782 & 0.00004 & 0.00651 \\ 
    \bottomrule
  \end{tabular}
  \label{tab:errors}
  \end{center}
\end{footnotesize}
\end{table}

\begin{figure}[h]
\centerline{\includegraphics[width=1\columnwidth]{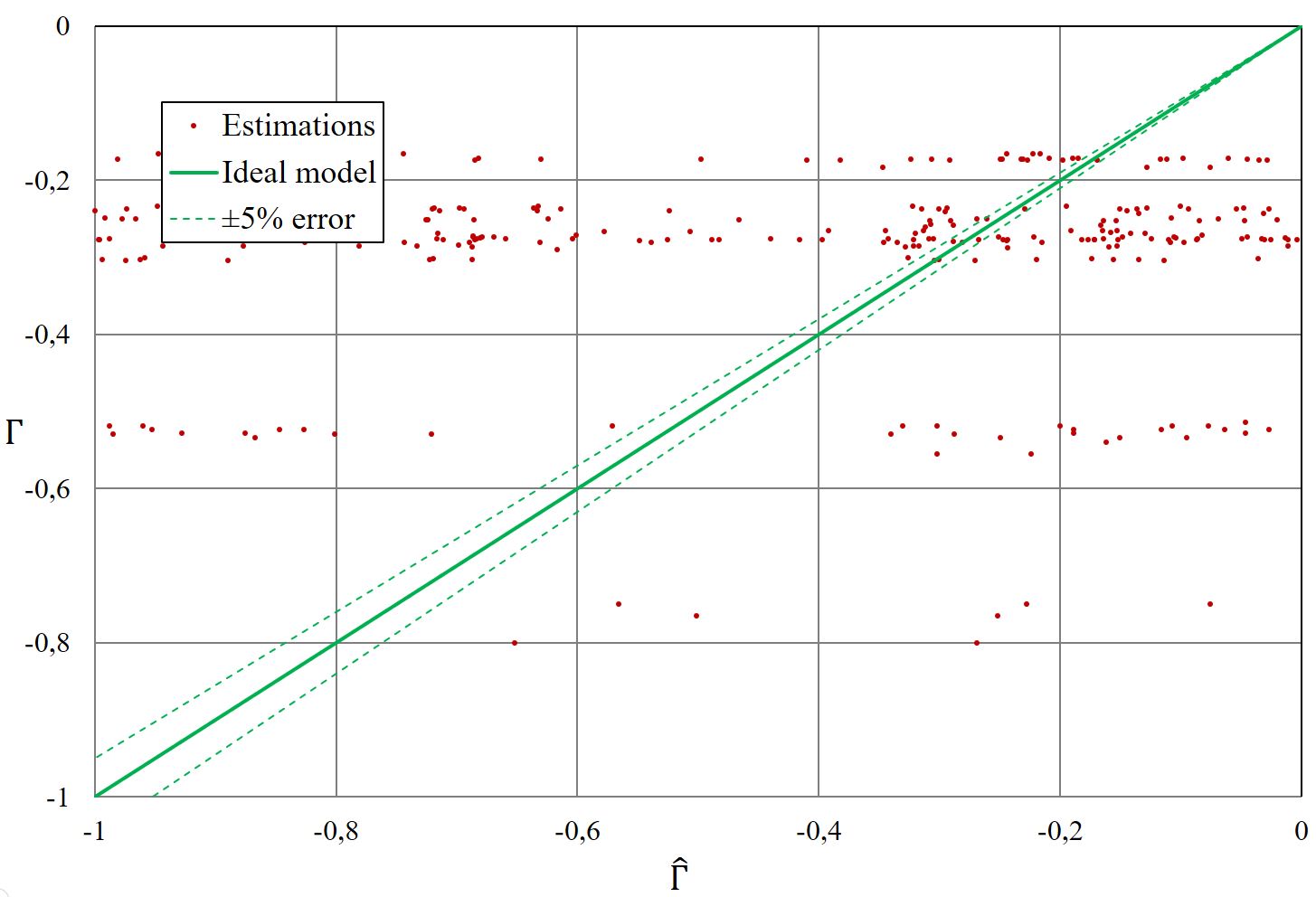}}
\caption{Estimated consumption $\hat{\Gamma}$ (red dots) of Model A with validation data set compared with the true consumption $\Gamma$, ideal model output (green line), and 5\% relative estimation error (green dashed lines).}
\label{fig:gamma_mlp}
%\vspace{-0.3cm}
\end{figure}

Showing the results graphically further emphasizes this finding:
Figure \ref{fig:gamma_mlp} illustrates the validation results of Model A. 
For each validation sample with consumption $\Gamma$ the estimation $\hat{\Gamma}$ that the model provided is depicted as red dot.
The green line represents an ideal model behavior for comparison.
The green dashed lines represent an RMAE of 5\%.
It can be seen, that most of the validation samples are not estimated correctly by the linear Model A, shown by the red dots that are far from the green line. 
Based on the low accuracy of the linear model, we assume the real relation between the input features $x_{1,..,90}$ and the consumption $\Gamma$ to be non-linear.

\begin{figure}[h]
\centerline{\includegraphics[width=1\columnwidth]{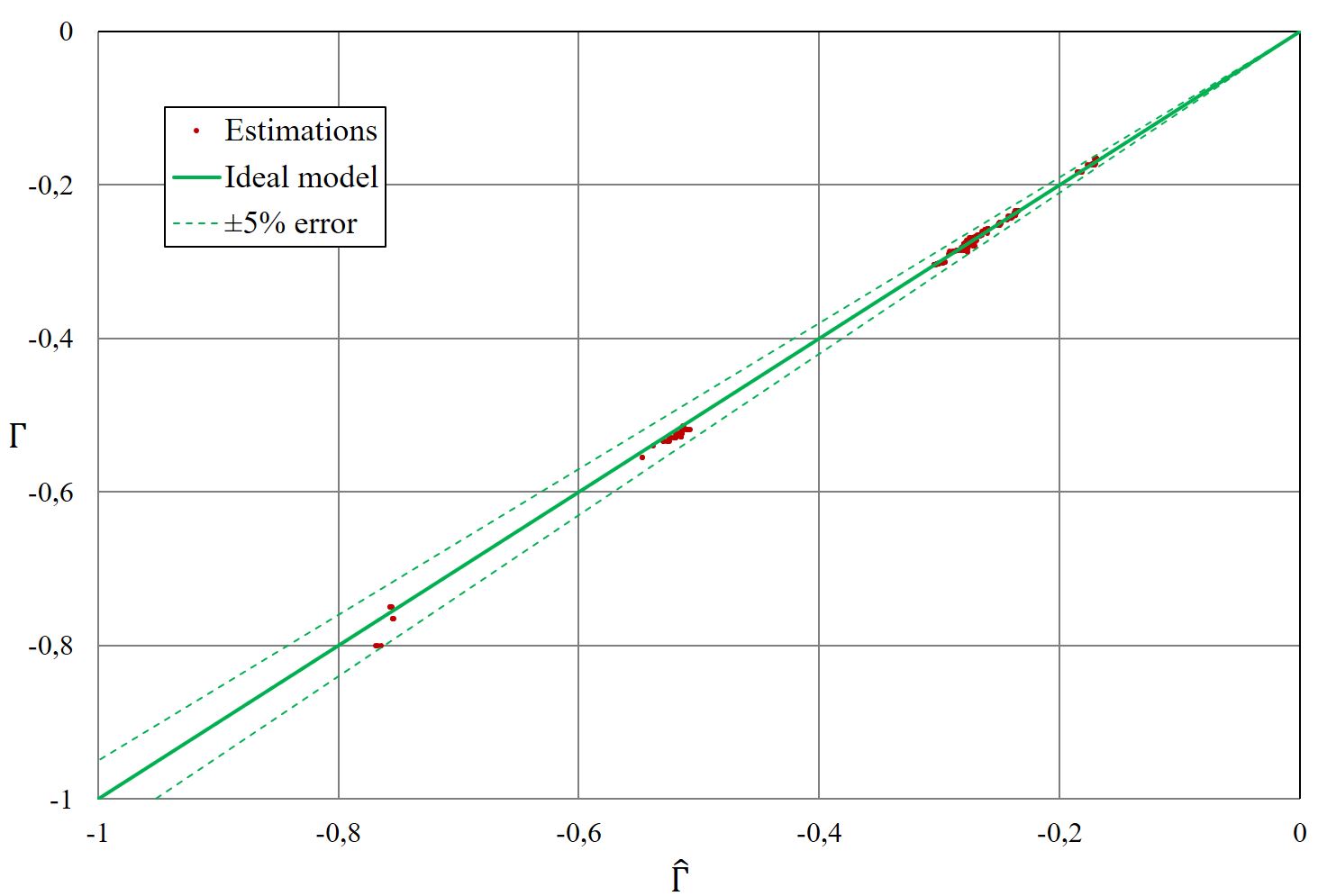}}
\caption{Estimated consumption $\hat{\Gamma}$ (red dots) of Model B with validation data set compared with the true consumption $\Gamma$, ideal model output (green line), and 5\% relative estimation error (green dashed lines).}
\label{fig:gamma_nn}
%\vspace{-0.3cm}
\end{figure}

The results of the neural network Model B support this assumption:
In the same manner as done for Model A, Fig. \ref{fig:gamma_nn} shows the estimations for Model B.
The results show a higher precision, as the estimations $\hat{\Gamma}$ are located closer to the ideal model representation, i.e. the green line.
A mean relative estimation error $\epsilon_\textrm{B} = 1.78$ \% (cf. Table \ref{tab:errors}) on the validation data supports this result.
According to the results, the internal structure of the neural network Model B seems to represent the influences on the energy consumption with much higher precision than the linear Model A.

A further improvement of the model performance is expected if further data, e.g. the actual weight of the electric vehicle due to baggage and passengers, would be considered.
Cross-validation over several vehicles should be proceeded to test the generalization of the developed model.

Note that the input data excludes signals allowing to infer the state of battery degradation.
Thus, the consumption model does not consider effects of battery aging.
Prospectively, battery aging can be estimated by training a similar model with data solely obtained from EVs, whose batteries have not degraded yet.
Then, a consumption can be estimated for vehicles that are assumed to have an aged battery.
A discrepancy between the estimation and the real consumption---beyond the known estimation error---indicates a battery aging due to increased internal losses.
This method would allow an entirely data-driven battery aging estimation based on data anyway created within the vehicle operation.
\pagebreak
\section{Conclusion and perspective} \label{sec:conclusion}
In this paper, we examined perspectives and interactions of Electric Vehicle (EV) users, car manufacturers and power suppliers.
We proposed a concept to quantify the objectives of all parties, aiming at automated calculation of EV charging strategies.
In this context, we outlined a cloud-based framework for data acquisition, storage and application as a common basis for data-driven analyses.
As an example use case, we proposed two data-driven models based on linear regression and neural-network regression to estimate specific consumption, i.e. consumed energy per driven distance.
Therefore, we used time series data collected from a fleet of ten connected EVs.
For the model training only eight out of ten EVs were used, while the remaining two EVs were used for validation.

The linear model seems to inadequately represent the true relation between input features and consumption.
However, the neural network model with five hidden layers yields a mean relative absolute error of 1.78\%.
Considering the underlying data, the proposed model outperforms estimators based on physical models as described in the literature.
More elaborate model validation with different data and other models, such as polynomial models, or other neural network architectures may allow further model precision improvement.

In future work, a model as proposed could estimate battery aging, if the training data is solely obtained from EVs, whose batteries have not degraded yet.
Comparing the estimated consumption with the actual consumption for EVs, which have degraded batteries, might indicate battery aging due to increased internal losses.
This method would allow an entirely data-driven battery aging estimation.
%Once the user behavior can be predicted with sufficient precision, the influence of single trips and charging phases on EV battery degradation might be inferred.

Furthermore, the remaining model parts of the proposed concept require a more elaborate implementation and evaluation.
The interactions among EV user, power supplier, and car manufacturer should be analyzed and described mathematically.
Finally, the findings need to be combined in an optimization approach, enabling the automated calculation of individualized EV charging strategies considering momentary grid load and battery preservation.
For concept evaluation, we plan to use an experimental EV fleet within measure campaigns in the \textit{Energy Lab 2.0} \cite{Hagenmeyer16}.
Concluding, the feasibility of grid-optimal, battery-preserving and individualized charging strategies needs to be investigated with adequate indexes, such as user acceptance.
%For instance, a conceivable metric for user acceptance could be the amount of time the user is actually following the stipulated charging plan.
%
%- focus on charging behavior and its influence on battery aging\\
%- conclude previous evaluations on battery health etc.\\
%- evaluation of customer, i.e. classification methods, behavior recognition, incentives? \\
%- influence on charging behavior, user acceptance?\\
%- how to cluster customers by charging behavior?\\
%- grid interaction, how can charging strategies be adapted etc.\\
%
%
%FURTHER QUESTIONS:
%
%\begin{enumerate}
%	\item How can customer acceptance of charging strategies be guaranteed?
%	\item How far can the contrary objectives be combined in charging strategies?
%\end{enumerate}
%
%\pagebreak
%
%\bibliographystyle{IEEEtran}
%\bibliography{literature/ciws}


%\includePaperPDF{paper_pdf}{\AuthorsTOC}{\AffiliationsTOC}{\Title}

\end{document}